\PassOptionsToPackage{numbers}{natbib}
\documentclass[10.5pt,compsoc,UTF8]{CjC}
\usepackage{CTEX}
\usepackage{graphicx}
\usepackage{footmisc}
\usepackage{subfigure}
\usepackage{url}
\usepackage{multirow}
\usepackage{multicol}
\usepackage[noadjust]{cite}
\usepackage{amsmath,amsthm}
\usepackage{amssymb,amsfonts}
\usepackage{booktabs}
\usepackage{color}
\usepackage{xcolor}
\usepackage{ccaption}
\usepackage{booktabs}
\usepackage{float}
\usepackage{fancyhdr}
\usepackage{caption}  
\usepackage{xcolor,stfloats}
\usepackage{comment}
\graphicspath{{figures/}}
\usepackage{cuted}
\usepackage{captionhack}
\usepackage{epstopdf}
\usepackage{gbt7714}
\usepackage{hyperref}  % 使用"\href"命令, 加载超链接
\usepackage{array, tabularx, multirow} % 加载大型表格(含有合并情况)
\usepackage{longtable} % 跨页续表
\usepackage{breqn} % 公式自动换行
\usepackage{placeins}

\newcommand{\deepPurple}[1]{\textcolor{purple}{\textbf{#1}}}
\newcommand{\lightPurple}[1]{\textcolor{violet}{\textbf{#1}}}
\newcommand{\ScaledSlash}[1]{\mathrel{\vcenter{\hbox{\scalebox{#1}{$/$}}}}}

\headevenname{\mbox{\quad} \hfill  \mbox{\zihao{-5}{ \hfill Hypergraph Mining via Proximity Matrix} }}
\headoddname{Mining Hypergraph via Proximity Matrix}

\newtheoremstyle{mystyle}{0pt}{0pt}{\normalfont}{1em}{\bf}{}{1em}{}
\theoremstyle{mystyle}
\renewcommand\figurename{figure~}

\makeatletter
\renewcommand{\@biblabel}[1]{[#1]\hfill}
\makeatother
\begin{document}
\hyphenpenalty=50000
\makeatletter
\newcommand\mysmall{\@setfontsize\mysmall{7}{9.5}}
\newenvironment{tablehere}
  {\def\@captype{table}}
  
\let\temp\footnote
\renewcommand \footnote[1]{\temp{\zihao{-5}#1}}
\newcolumntype{C}{>{\centering\arraybackslash}X}  
\thispagestyle{plain}
\thispagestyle{empty}
\pagestyle{CjCheadings}

\vspace {-13mm}
{
\centering
\vspace {12mm}
{\zihao{2} \heiti Hypergraph Mining via Proximity Matrix} 

\vspace{10mm}

{\zihao{4} 
Junhao Bian$^{1}$, Yilin Bi$^{1}$, Tao Zhou$^{1,*}$}

\vspace{2mm}

\begin{center}
\zihao{5}{$^{1}$CompleX Lab, School of Computer Science and Engineering, 
University of Electronic Science and Technology of China, Chengdu 610054, China}\

\zihao{5}{$^{*}$Corresponding author: \href{zhutou@ustc.edu}{zhutou@ustc.edu}}
\end{center}
}

\vskip 5mm
\zihao{5}
\setlength{\baselineskip}{16pt}\selectfont

\begin{center}
\textbf{ABSTRACT}
\end{center}

\vspace{2mm}
\noindent
Hypergraphs serve as an effective tool widely adopted to characterize higher-order interactions in complex systems. The most intuitive and commonly used mathematical instrument for representing a hypergraph is the incidence matrix, in which each entry is binary, indicating whether the corresponding node belongs to the corresponding hyperedge. Although the incidence matrix has become a foundational tool for hypergraph analysis and mining, we argue that its binary nature is insufficient to accurately capture the complexity of node-hyperedge relationships arising from the fact that different hyperedges can contain vastly different numbers of nodes. Accordingly, based on the resource allocation process on hypergraphs, we propose a continuous-valued matrix to quantify the proximity between nodes and hyperedges. To verify the effectiveness of the proposed proximity matrix, we investigate three important tasks in hypergraph mining: link prediction, vital nodes identification, and community detection. Experimental results on numerous real-world hypergraphs show that simply designed algorithms centered on the proximity matrix significantly outperform benchmark algorithms across these three tasks.

\vspace{2mm}
\noindent\textbf{Keywords:} Hypergraph mining, Proximity matrix, Resource allocation, Link prediction, Vital nodes identification, Community detection

\section{Introduction}

The complexity in biological, social, economic, and technological systems largely stems from the interactions among entities \cite{kauffman1996, barabasi2007, hidalgo2018}. Networks, as a powerful mathematical tool, have been widely adopted to model these interactions by representing entities as nodes and pairwise relationships as links \cite{barabasi2016, newman2018}. However, traditional network models fall short in capturing interactions involving multiple entities. Since simplifying multi-body interactions into multiple pairwise interactions will result in significant information loss, using hypergraphs to characterize such multi-body interactions becomes an inevitable choice \cite{yoon2020, bian2025}. Accordingly, research on hypergraphs has now become the most productive branch in network science \cite{battiston2020, battiston2021, battiston2022, bianconi2021, bick2023}.

Graph mining is an interdisciplinary technique that extracts hidden patterns, important entities, correlation rules, abnormal features, and other valuable information from network structures \cite{rehman2012}. In complex network research, three typical graph mining tasks--link prediction \cite{lu2011, zhou2021}, vital nodes identification \cite{lu2016, chen2025}, and community detection \cite{newman2006, fortunato2010}--have received widespread attention and found applications in fields such as drug-target prediction \cite{csermely2013}, personalized recommendation \cite{lu2012}, inference of network evolution mechanisms \cite{wang2012}, infectious disease prevention \cite{chaharborj2022}, critical gene identification \cite{bi2022}, and discovery of research hotspots \cite{tang2008}. Similar to graph mining, with the emergence of hypergraph research, hypergraph mining has also become a hot topic \cite{antelmi2023, tian2024, lee2025}, and an increasing number of scientists have started to investigate link prediction \cite{zhang2018, benson2018, kumar2020, contisciani2022, chen2023, deng2025}, vital nodes identification \cite{zhu2018, tudisco2021, antelmi2021, hu2023, xie2023a, xie2023b, zeng2024, zhang2024, sun2025}, and community detection \cite{li2017, kaminski2019, ahn2019, kumar2020b, yuan2022, chodrow2023, ruggeri2023, kovacs2025} on hypergraphs.

The matrix representation of graphs plays a crucial role in graph mining, as these matrices can fully preserve graph information and offer convenience for mathematical derivation and algorithm design. Among them, the two most widely used matrices are the adjacency matrix $\mathcal{A}$ and the Laplacian matrix $\mathcal{L}$. Many graph mining algorithms are based on them, such as non-negative matrix factorization \cite{chen2017} and low-rank matrix completion  \cite{pech2017} in link prediction, eigenvector centrality \cite{bonacich1987} and subgraph centrality \cite{estrada2005} in vital nodes identification, and spectral methods \cite{newman2013} and singular value decomposition \cite{sarkar2011} in community detection, to name just a few. Unfortunately, we cannot directly characterize hypergraphs using adjacency matrices or Laplacian matrices \cite{chung1993,nurisso2025}. The reason is that the relationship between two nodes in a hypergraph is very complex: they may not belong to any hyperedge simultaneously, or they may be contained many hyperedges at the same time. Furthermore, the tightness between two nodes differs when they are included in a hyperedge with only a few nodes compared to when they are in a hyperedge with many nodes. A natural idea is to use the incidence matrix $\mathcal{M}$ as an alternative. $\mathcal{M}$ is a binary matrix, where $\mathcal{M}_{i\alpha}=1$ if the node $v_i$ belongs to the hyperedge $e_\alpha$ and $\mathcal{M}_{i\alpha}=0$ otherwise. Obviously, the incidence matrix fully preserves the information of a hypergraph. However, the orders of hyperedges (an order of a hyperedge is defined as the number of nodes contained in this hyperedge) in real-world hypergraphs exhibit a highly heterogeneous distribution \cite{kook2020}. Therefore, the significance of a node belonging to a very small hyperedge is quite different from that of belonging to a very large one, but such a difference is indistinguishable in the incidence matrix. As can be seen from numerous comparative experiments later in this paper, the direct use of the incidence matrix does not yield sufficiently good performance in hypergraph mining.

We believe that the performance of hypergraph mining algorithms can be improved by constructing a matrix that better quantifies the proximity between nodes and hyperedges in hypergraphs. In this paper, we build the aforementioned proximity matrix by leveraging the resource allocation process on hypergraphs. We conducted extensive experiments on the three important hypergraph mining tasks (i.e., link prediction, vital nodes identification, and community detection) and found that the algorithms based on the proximity matrix outperform state-of-the-art algorithms in all three tasks. The proximity matrix we constructed can serve as a fundamental tool for hypergraph mining and hypergraph analysis, and its applications are not limited to the three tasks considered in this paper.

\section{Proximity Matrix}

A hypergraph is denoted as $G(V, E)$, where $V = \{v_1, v_2,\dots, v_N \}$ is the set of nodes, and $E = \{e_1, e_2,\dots, e_M \}$ is the set of hyperedges \cite{berge1984, bretto2013}. The number of hyperedges containing node $v_{i}$ is defined as the degree of $v_{i}$, denoted as $d_i$. Unlike traditional edge that represents a connection between two nodes, a hyperedge can capture the relationship among multiple nodes. Specifically, a hyperedge $e_\alpha$ can contain two or more nodes, and its order, denoted by $k_{\alpha}$, is defined as the number of nodes belonging to $e_\alpha$. 

\begin{figure}[t] 
        \centering \includegraphics[width=0.85\columnwidth]{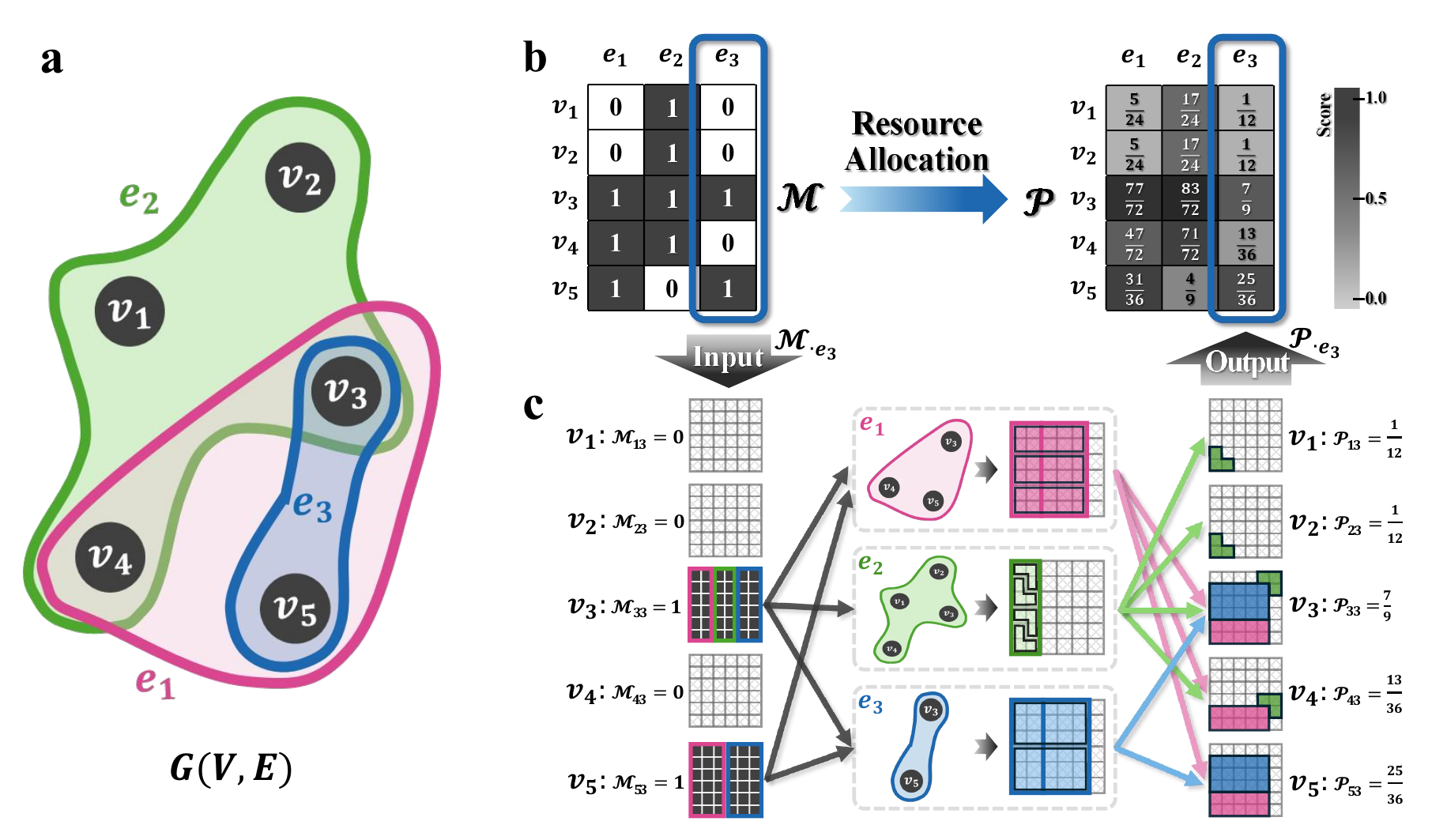}
        \caption{\label{f_HRA} Illustration of hypergraph resource allocation. (a) An example hypergraph. (b) The initial matrix $\mathcal{M}$ and the resulted matrix $\mathcal{P}$. (c) The details of the two steps of resource allocation to obtain $\mathcal{P}$ from $\mathcal{M}$, with only the vector associated with hyperedge $e_3$ is presented. 
        }
\end{figure}

As mentioned earlier, merely using a binary incidence matrix is insufficient to characterize the proximity between nodes and hyperedges. Therefore, we draw on the idea of the resource allocation process in networks \cite{ou2007,zhou2007,zhou2009,zhou2010} to quantify such proximity. We consider a conservative resource allocation process, where each round of allocation consists of two steps: (1) The resources on each node $v_i$ are equally divided into $d_i$ parts, which are then allocated to its adjacent hyperedges; (2) The resources on each hyperedge $e_\alpha$ are equally divided into $k_\alpha$ parts, which are then allocated to the nodes contained in this hyperedge. Accordingly, we employ a Markov transition matrix $\mathcal{T}$ to characterize the above process, say 
\begin{equation}
\mathcal{T} =  (\mathcal{M} \cdot \mathcal{K}^{-1}) \cdot (\mathcal{M}^\top \cdot \mathcal{D}^{-1}),
\end{equation}
where $\mathcal{K}$ ($M \times M$) and $\mathcal{D}$ ($N \times N$) are two diagonal matrices defined as $\mathcal{K}_{\alpha \alpha}=k_\alpha$ and $\mathcal{D}_{ii}=d_i$, respectively. The initial resource distribution is naturally set to be $\mathcal{M}$, where  $\mathcal{M}_{i\alpha}$ denotes the resource of nodes $v_i$ contributed by hyperedge $e_\alpha$. Then, the proximity matrix $\mathcal{P}$, obtained by the resource allocation process, can be formulated as 
\begin{equation}
\mathcal{P} =  \mathcal{T} \cdot \mathcal{M}.
\end{equation}
Figure~\ref{f_HRA} illustrates the resource allocation process, say $\mathcal{M} \rightarrow \mathcal{P}$, on an example hypergraph. In particular, Figure~\ref{f_HRA}c shows the detailed procedure associated with hyperedge $e_3$. 

\section{Results}

Although the method to obtain $\mathcal{P}$ is straightforward, such a matrix can be successfully applied to multiple aspects of hypergraph analysis. Below, we mainly focus on the applications of $\mathcal{P}$ in three important tasks of hypergraph mining: link prediction, community detection, and vital nodes identification.

\begin{table}[ht]
\centering
\caption{Fundamental statistics of the 18 hypergraphs under consideration, including the number of nodes $N$, the number of hyperedges $M$, the average degree $\langle d \rangle$, the average hyperedge order $\langle k \rangle$, and the number of labeled communities $N_c$.}
\label{DS}
\begin{tabular}{lccccccc}
\toprule
\textbf{Hypergraphs} & \textbf{$N$} & \textbf{$M$} & \textbf{$\langle d \rangle$} & \textbf{$\langle k \rangle$} & \textbf{$N_c$} & \textbf{References}\\
\midrule
email-Enron & 143 & 1487 & 32.12 & 3.09 & N/A & \cite{benson2018} \\
email-Eu & 564 & 1142 & 5.61 & 2.77 & N/A & \cite{benson2018, Yin-2017, Leskovec-2007} \\
DAWN & 897 & 7745 & 27.34 & 3.17 & N/A & \cite{benson2018} \\
NDC-classes & 1149 & 1181 & 6.75 &  6.56 & N/A & \cite{benson2018} \\
congress-bills & 534 & 3738 & 71.19 &  10.17 & N/A & \cite{benson2018, Fowler-2006} \\
contact-high-school & 312 & 2669 & 19.31 &  2.26 & N/A & \cite{benson2018, Mastrandrea-2015} \\
contact-primary-school & 231 & 2722 & 26.42 &  2.24 & N/A & \cite{benson2018, Mastrandrea-2015} \\
tags-ask-ubuntu & 702 & 1424 & 5.82 &  2.87 & N/A & \cite{benson2018} \\
tags-math-sx & 406 & 1141 & 7.71 &  2.74 & N/A & \cite{benson2018} \\
iAF1260b & 1666 & 2022 & 5.00 &  4.12 & N/A & \cite{King-2015} \\
iAF987 & 1098 & 1163 & 4.74 &  4.48 & N/A & \cite{King-2015} \\
iCN900 & 883 & 1090 & 5.57 &  4.51 & N/A & \cite{King-2015} \\
Cora & 2708  & 1429  & 18.16 &  34.41 & 7 & \cite{sen2008} \\
Citeseer & 3312  & 3625  & 31.61 &  28.88 & 6 & \cite{sen2008} \\
High-school & 327   & 7818  & 55.63 &  2.33  & 9 & \cite{Mastrandrea-2015, chodrow2021} \\
Primary-school & 242   & 12704 & 126.98 &  2.42  & 11 & \cite{chodrow2021, stehle2011} \\
Senate-committees & 282   & 302   & 18.85 &  17.60 & 2 & \cite{chodrow2021, stewart2017S} \\
House-committees & 1290  & 335   & 9.17 &  35.31 & 2 & \cite{chodrow2021, stewart2017H} \\
\bottomrule
\end{tabular}
\end{table}

To evaluate the performance of the three above-mentioned tasks, we utilize 18 real-world hypergraphs from diverse domains. A brief introduction of these hypergraphs is shown below. \textbf{email-Enron, email-Eu:} These two hypergraphs capture internal email communication within the Enron corporation \cite{benson2018} and the European research institution \cite{benson2018, Yin-2017, Leskovec-2007}. Nodes represent employee email addresses, and each hyperedge consists of the sender and all recipients of a single email. \textbf{DAWN:} This hypergraph originates from the Drug Abuse Warning Network (DAWN) in the United States \cite{benson2018}, recording nationwide emergency department visits related to drug use, where nodes are unique drug IDs, and hyperedges comprise all drugs used in the same emergency incident. \textbf{NDC-classes:} This is based on the National Drug Code (NDC) directory of the United States \cite{benson2018}, where nodes denote drug class labels, and each hyperedge represents multiple class labels involved in one drug. \textbf{congress-bills:} This hypergraph is constructed from the U.S. Congress bills data set \cite{benson2018, Fowler-2006}, where nodes represent members of Congress, and each hyperedge contains all sponsors and co-sponsors of a bill. \textbf{contact-high-school, contact-primary-school:} These two hypergraphs capture face-to-face interactions in high-school and primary-school environments \cite{benson2018, Mastrandrea-2015}. Nodes represent students and staff, while each hyperedge consists of all individuals engaged in physical contact within a 20-second time window, thereby reflecting short-term group interactions. \textbf{tags-ask-ubuntu, tags-math-sx:} These are two tag co-occurrence hypergraphs from the Ask Ubuntu and Math Stack Exchange platforms \cite{benson2018}, where nodes represent technical tags and each hyperedge corresponds to a question together with all its associated tags, reflecting the collaborative multi-label annotation pattern. \textbf{iAF1260b, iAF987, iCN900:} These hypergraphs are derived from the Biochemical, Genetic, and Genomic (BiGG) metabolic networks \cite{King-2015}. Each metabolite is modeled as a node, and each biochemical reaction forms a hyperedge, connecting the participating reactant and product metabolites. \textbf{Cora, Citeseer:} These hypergraphs originate from citation networks \cite{sen2008}. Nodes correspond to academic papers, and each hyperedge consists of all keywords associated with a paper. Community labels are the disciplinary categories of papers. \textbf{High-school, Primary-school:} The extended school contact networks \cite{Mastrandrea-2015, chodrow2021, stehle2011}, where nodes include both students and teachers, and community labels are grade levels. \textbf{Senate-committees, House-committees:} These are U.S. Senate and House committee membership hypergraphs \cite{chodrow2021, stewart2017S, stewart2017H}, where nodes represent legislators and hyperedges correspond to standing committees, with community labels indicating party affiliations (Democrat or Republican). 

Among these hypergraphs, \textbf{Cora}, \textbf{Citeseer}, \textbf{High-school}, \textbf{Primary-school}, \textbf{Senate-committees}, and \textbf{House-committees} have community labels (i.e., the ground-truth of community detection), and are thus specifically used to evaluate the performance of community detection. The remaining 12 hypergraphs are used for link prediction and vital nodes identification. Table~\ref{DS} presents fundamental statistics of all 18 hypergraphs. 

\subsection{Link Prediction}
The link prediction task in hypergraphs aims to predict which sets of nodes will form hyperedges based on the observed structure. These hyperedges may either not yet exist but emerge as the hypergraph evolves, or already exist but  unobserved \cite{chen2023}. To evaluate the prediction accuracy, hyperedges in the target hypergraph are randomly split into a training set and a test set, where the former is utilized to train the model, and the latter is used to evaluate the performance. This study employs 5-fold cross-validation, where the set of hyperedges is randomly and evenly divided into 5 subsets. One subset is used as the test set, and the remaining four subsets serve as the training set. The entire evaluation process is repeated 5 times, with a different test set selected each time, and the final result is the average value over the 5 runs. In link prediction for simple networks, all unobserved links can be treated as negative samples \cite{lu2011}. However, this approach is computationally infeasible for hypergraphs, because the number of unobserved hyperedges scales exponentially ($O(2^N)$). At the same time, the random sampling with negatives being uniformly drawn from all unobserved hyperedges is also ineffective because such negatives are too easy to be distinguished from positives, and thus provide less value in training a good link prediction model \cite{patil2020,hwang2022,deng2025}. We propose a negative sampling method that can adjust the similarity between negative and positive samples. Specifically, for each positive sample $e_\alpha$, we generate a corresponding negative sample $e^-_\alpha$, which contains the same number of nodes as the former, say $k_\alpha$. We randomly select $\rho k_\alpha$ nodes (if $\rho k_\alpha$ is not an integer, we set it to $\lfloor \rho k_\alpha \rfloor$ with probability $\rho k_\alpha - \lfloor \rho k_\alpha \rfloor$, and to $\lfloor \rho k_\alpha \rfloor +1$ with probability $1+\lfloor \rho k_\alpha \rfloor-\rho k_\alpha$， where $\lfloor x \rfloor$ denotes the floor of $x$) from $e_\alpha$ to serve as nodes in $e^-_\alpha$, and the remaining $(1-\rho) k_\alpha$ nodes in $e^-_\alpha$ are randomly sampled from the node in set $V$ that do not belong to $e_\alpha$. The larger $\rho$ corresponds to a higher similarity between negative and positive samples.

In principle, a hypergraph link prediction algorithm scores any set of nodes by the likelihood of forming a hyperedge; a higher score indicates a greater probability of the set being a hyperedge. As described in the previous section, the matrix $\mathcal{P}$ quantifies the proximity between nodes and hyperedges. Based on this, we define a similarity matrix $\mathcal{S}$ that measures the similarity between nodes, as:
\begin{equation}
\label{eq3}
\mathcal{S}_{ij} =
\begin{cases} 
\frac{\sum^N_{\alpha=1}\mathcal{P}_{i\alpha}\mathcal{P}_{j\alpha}}{\sqrt{d_id_j}}, & \text{if } i \neq j \\
0. & \text{otherwise}
\end{cases}
\end{equation}
Consistent with conventional methods for hypergraph link prediction, the score of a node set is set as the average value of the similarity between all node pairs in this set \cite{chen2023}. This method is named as Hypergraph Resource Allocation at Link Prediction (HRA@LP for short).

Taking into account the discriminative ability and complementarity of different metrics \cite{zhou2023,jiao2024,bi2024,wan2025}, this study adopts two metrics, area under the receiver operating characteristic curve (AUC) \cite{hanley1982,bradley1997} and normalized discounted cumulative gain (NDCG) \cite{jarvelin2002,wang2013}, to evaluate the performance of prediction algorithms. AUC measures the ability of the model to discriminate between positive and negative classes across all possible thresholds, which is equivalent to the probability that a randomly selected positive sample (i.e., a hyperedge in the test set) is scored higher than a randomly selected negative sample. If we randomly compare $n$ positive-negative pairs, and there are $n_1$ times the positive sample having higher score and $n_2$ times the two sharing the same score, then AUC is approximated as
\begin{equation}
\mathrm{AUC} = \frac{n_1+0.5n_2}{n}.
\end{equation} 
AUC will approach 0.5 if the algorithm scores node sets completely randomly, and thus to what extent AUC exceeds 0.5 indicates how much better the algorithm performs than pure chance. NDCG assigns larger weights to higher positions, normalized by the ideal discounted cumulative gain, as
\begin{equation}
\mathrm{NDCG}=
\left(\sum_{i=1}^{|E^P|}\frac{1}{\log_2(1+r_i)}\right)
\;\ScaledSlash{4.5}\;
\left(\sum_{l=1}^{|E^P|}\frac{1}{\log_2(1+l)}\right).
\end{equation}
where $E^P$ is the test set (i.e., the set of all positive samples), and $r_i$ is the rank of the $i$-th positive sample among all positive and negative samples (all samples are ranked in a descending order of their scores, and thus the sample ranked No. 1 is with the highest score). The value of NDCG lies in the interval $[0,1]$, where $\mathrm{NDCG}=1$ corresponds to a perfect ranking (i.e., all positive samples are placed ahead of negative samples).

\begin{table}[htbp]
\centering
\scriptsize 
\caption{Comparative results of the performance of algorithms, where red indicates the best performance under the corresponding evaluation metric and purple denotes the second-best. The parameter is set as $\rho=0.5$.}
\label{rho0.5}
\begin{tabular}{l
    |c|c 
    |c|c 
    |c|c  
    |c|c  
    |c|c  
}
\toprule
\textbf{Dataset}
  & \multicolumn{2}{c|}{\textbf{CN}}
  & \multicolumn{2}{c|}{\textbf{HPRA}}
  & \multicolumn{2}{c|}{\textbf{Katz}}
  & \multicolumn{2}{c|}{\textbf{NHNE}} 
  & \multicolumn{2}{c}{\textbf{HRA@LP}} \\
& \textbf{AUC} & \textbf{NDCG}
& \textbf{AUC} & \textbf{NDCG}
& \textbf{AUC} & \textbf{NDCG}
& \textbf{AUC} & \textbf{NDCG}
& \textbf{AUC} & \textbf{NDCG} \\
\midrule
\textbf{email-Enron} & 0.8067 & 0.9544 & {0.8683} & \lightPurple{0.9749} & 0.8267 & 0.9584 & \lightPurple{0.8855} & 0.9683 & \deepPurple{0.9024} & \deepPurple{0.9863} \\
\textbf{email-Eu} & 0.7698 & 0.9621 & {0.7877} & \lightPurple{0.9690} & 0.7729 & 0.9669 & \lightPurple{0.7960} & 0.9471 & \deepPurple{0.8140} & \deepPurple{0.9738} \\
\textbf{DAWN} & 0.5910 & 0.9357 & {0.8272} & \lightPurple{0.9766} & 0.5735 & 0.9349 & \lightPurple{0.8672} & 0.9561 & \deepPurple{0.8787} & \deepPurple{0.9857} \\
\textbf{NDC-classes} & 0.7014 & 0.9603 & 0.7083 & 0.9544 & {0.7975} & \deepPurple{0.9805} & \deepPurple{0.8954} & 0.9800 & \lightPurple{0.7960} & \lightPurple{0.9786} \\
\textbf{congress-bills} & 0.6615 & 0.9430 & \lightPurple{0.7978} & \lightPurple{0.9701} & 0.6809 & 0.9444 & 0.6434 & 0.8827 & \deepPurple{0.8230} & \deepPurple{0.9745} \\
\textbf{contact-high-school} & {0.9072} & \lightPurple{0.9865} & 0.9067 & 0.9853 & \lightPurple{0.9126} & 0.9855 & 0.8329 & {0.9588} & \deepPurple{0.9373} & \deepPurple{0.9917} \\
\textbf{contact-primary-school} & \lightPurple{0.9286} & \lightPurple{0.9816} & 0.9253 & 0.9788 & 0.9122 & 0.9779 & 0.9276 & 0.9797 & \deepPurple{0.9624} & \deepPurple{0.9941} \\
\textbf{tags-ask-ubuntu} & 0.7741 & 0.9647 & {0.7793} & \lightPurple{0.9697} & 0.7770 & 0.9683 & \lightPurple{0.7902} & 0.9483 & \deepPurple{0.7933} & \deepPurple{0.9709} \\
\textbf{tags-math-sx} & 0.8297 & {0.9739} & \deepPurple{0.8562} & \deepPurple{0.9795} & 0.8451 & \lightPurple{0.9771} & 0.8374 & 0.9453 & \lightPurple{0.8457} & 0.9723 \\
\textbf{iAF1260b} & 0.7041 & 0.9594 & {0.7158} & {0.9620} & {0.7392} & {0.9668} & \deepPurple{0.8482} & \lightPurple{0.9697} & \lightPurple{0.7560} & \deepPurple{0.9737} \\
\textbf{iAF987} & 0.7315 & 0.9624 & {0.7537} & {0.9651} & 0.7528 & 0.9620 & \deepPurple{0.8445} & \lightPurple{0.9685} & \lightPurple{0.8043} & \deepPurple{0.9761} \\
\textbf{iCN900} & 0.7786 & 0.9683 & {0.7957} & {0.9691} & {0.8081} & {0.9707} & \deepPurple{0.8763} & \lightPurple{0.9759} & \lightPurple{0.8462} & \deepPurple{0.9805} \\
\bottomrule
\end{tabular}
\end{table}

We consider the following four benchmark hypergraph link prediction algorithms (see Methods for algorithm details):  common neighbors index (CN) \cite{liben2007, zhang2018} based on local neighborhood, hyperedge prediction using resource allocation (HPRA)  \cite{kumar2020}, Katz index (Katz) \cite{katz1953, zhang2018} based on global paths, and nonuniform hypergraph embedding with dual mechanism (NHNE) \cite{huang2020}. Table~\ref{rho0.5} compares the performance of the HRA@LP method with those four benchmark methods in twelve real-world hypergraphs, measured by AUC and NDCG. As shown in Table~\ref{rho0.5}, HRA@LP exhibits remarkable advantage for both AUC and NDCG. Table S1 (see Section S1 in SI Appendix) provides the comparative results of the performance of the five algorithms under more $\rho$-values ($\rho=0.6$, $\rho=0.7$, and $\rho=0.8$). From these results, it can be observed that the HRA@LP method consistenly exhibits stable and significant advantages.

\subsection{Community Detection}
Community detection generally requires partitioning all nodes into several communities (\textit{i.e.}, clusters), where the intra-community connections are dense and the inter-community connections are relatively sparse \cite{newman2006,  fortunato2010}. The spectrum of a graph is commonly used to address the community detection problem \cite{chung1997, newman2006b, newman2013}. By transforming the problem of community detection into the minimum cut problem of a graph, researchers have proposed performing spectral decomposition on the symmetric normalized Laplacian matrix, and then determining the community each node belongs to based on the node's components in the eigenvectors \cite{shi2000, newman2006b}.One can then represent each node as an $N_c$-dimensional vector consisting of its components in the eigenvectors for the first $N_c$ smallest eigenvalues, then cluster these real vectors with well-established clustering methods. This approach is also applicable to hypergraph community detection \cite{zhouD2006}. Inspired by the above studies, we propose a spectral method for hypergraph community detection. Specifically, we first construct a symmetric normalized Laplacian matrix
\begin{equation}
  \mathcal{L} = \mathcal{I} - \mathcal{D}^{-\frac{1}{2}} \mathcal{S} \mathcal{D}^{-\frac{1}{2}},
\end{equation}
where the matrix $\mathcal{S}$ is the similarity matrix derived from the proximity matrix $\mathcal{P}$ (see Eq.~\ref{eq3}). Next, we represent each node as an $N_c$-dimensional vector consisting of its components in the eigenvectors for the first $N_c$ smallest eigenvalues of $\mathcal{L}$ and directly apply the famous $k$-means (here, $k=N_c$) algorithm \cite{macqueen1967, hartigan1979} to get the partition of communities. Similar to the scenario for link prediction, this method is named as Hypergraph Resource Allocation at Community Detection (HRA@CD for short).

To evaluate the algorithm performance, we conduct experiments on the six hypergraphs with ground-truth community labels (Cora, Citeseer, High-school, Primary-school, Senate-committees, and House-committees, see Table~\ref{DS}) and adopts Precision as the metric. Given a community detection algorithm, for every pair of nodes $(v_i,v_j)$ ($i\neq j$) that are assigned to the same community by the algorithm, this node pair is referred to as a \textbf{hit} if the two nodes actually belong to the same community based on the ground-truth community labels; conversely, it is termed a \textbf{miss}. Accordingly, the Precision is defined as the hitting rate of all node pairs that are assigned to the same community by the algorithm. We compare the proposed HRA@CD method with the following five benchmark hypergraph community detection algorithms (see Methods for algorithm details): non-negative matrix factorization (NMF) \cite{lee2000}, node degree preserving Louvian algorithm (NDP-Louvian) \cite{kumar2020b}, Metis algorithm for hypergraph adjacency matrix (AMetis) \cite{karypis1998, catalyurek1999}, hypergraph spectral clustering (HSC) \cite{zhouD2006}, and nonbacktracking hypergraph spectral clustering (NBHSC) \cite{chodrow2023}. As shown in Table~\ref{CD}, HRA@CD performs notably better than those benchmarks. In particular, HRA@CD shows remarkably prominent advantages on Cora and Citeseer, and the community partitioning problem for these two hypergraphs is precisely the most challenging. This is because they feature not only a considerably large average order $\langle k \rangle$ of hyperedges but also a relatively large number of communities $N_c$. The success of HRA@CD for those difficult cases indicates that the proximity matrix $\mathcal{P}$ we constructed and the corresponding similarity matrix $\mathcal{S}$ can appropriately characterize the proximity (for node-hyperedge pairs) and similarity (for node-node pairs) resulted from high-order interactions.

\begin{table}[htbp]
\centering
\caption{Comparative results of the performance of community detection algorithms, where red indicates the best performance measured by Precision and purple denotes the second-best.}
\label{CD}
\begin{tabular}{l|c|c|c|c|c|c}
\toprule
\textbf{Datasets} & \textbf{NMF} & \textbf{NDP-Louvian} & \textbf{AMetis} & \textbf{HSC} & \textbf{NBHSC} & \textbf{HRA@CD} \\
\midrule
\textbf{Cora} & 0.2990 & \lightPurple{0.3053} & 0.1786 & 0.2888 & 0.2064 & \deepPurple{0.3780} \\
\textbf{Citeseer} & \lightPurple{0.4017} & 0.3221 & 0.1781 & 0.3319 & 0.2162 & \deepPurple{0.5028} \\
\textbf{High-school} & 0.6782 & 0.7307 & 0.9877 & \deepPurple{0.9942} & 0.5589 & \lightPurple{0.9935} \\
\textbf{Primary-school} & 0.7285 & 0.5478 & 0.9000 & \lightPurple{0.9078} & 0.7746 & \deepPurple{0.9106} \\
\textbf{Senate-committees} & 0.4971 & 0.4965 & \lightPurple{0.4980} & \lightPurple{0.4980} & 0.4971 & \deepPurple{0.4988} \\
\textbf{House-committees} & 0.5001 & 0.5004 & 0.5003 & \lightPurple{0.5008} & 0.5003 & \deepPurple{0.5019} \\
\bottomrule
\end{tabular}
\end{table}

\subsection{Vital Nodes Identification (HRA@VNI)}

Vital nodes identification on hypergraphs aims to characterize influences of nodes, which are often reflected by the capacities of nodes to facilitate the spread of information, innovations, rumors, infections, and other contents on the hypergraph. To be more concrete, we take the hypergraph of scientific collaboration as an example, where nodes represent scientists, and scientists who co-author a paper form a hyperedge \cite{patania2017,juul2024,battiston2025}. If two hyperedges share a common scientist, scientific knowledge or innovative ideas are likely to spread between two research groups through this scientist. Therefore, if we denote this scientist by node $v_i$, and the two hyperedges by $e_\alpha$ and $e_\beta$, then it is natural to use the proximity between $v_i$ and $e_\alpha$ multiplied by that between $v_i$ and $e_\beta$ (i.e., $\mathcal{P}_{i\alpha}\mathcal{P}_{i\beta}$) to measure the possibility of the  exchange of scientific knowledge or innovative ideas through $v_i$. Accordingly, the centrality (influence) of node $v_i$ can be defined as
\begin{equation}
c_i^{\mathrm{HRA}}=\sum_{\alpha\neq\beta}\mathcal{P}_{i\alpha}\mathcal{P}_{i\beta}.
\label{eq7}
\end{equation}
In the calculation, we can use the quadratic form of Eq.~\ref{eq7}
\begin{equation}
c_i^{\mathrm{HRA}}=
\Bigl(\sum_{\alpha}\mathcal{P}_{i\alpha}\Bigr)^2 - \sum_{\alpha}\mathcal{P}_{i\alpha}^2.
\label{eq8}
\end{equation}
Since Eq.~\ref{eq8} depends only on the row sums and sum of squares of $\mathcal{P}$, its time complexity is only $O(\|\mathcal{P}\|_0)$, where $\|\mathcal{P}\|_0$ is the $L_0$ norm of $\mathcal{P}$, i.e., the number of nonzero entries of $\mathcal{P}$. We can sort the nodes according to $c_i^{\mathrm{HRA}}$: the larger the value of $c_i^{\mathrm{HRA}}$, the higher the node ranks, namely the more vital it is. This method is named as Hypergraph Resource Allocation at Vital Nodes Identification (HRA@VNI).

\begin{table}[htbp]
\centering
\caption{Comparative results of the performance of methods to identify vital nodes when $\beta = \beta_c$, where red indicates the best performance measured by the Kendall's $\tau$, and purple denotes the second-best.}
\label{VNI}
{\footnotesize
\setlength{\tabcolsep}{4pt}
\renewcommand{\arraystretch}{1.15}
\begin{tabular}{l|c|c|c|c|c|c}
\toprule
\textbf{Datasets} & \textbf{HEC} & \textbf{Katz} & \textbf{NB} & \textbf{SHC} & \textbf{HDC} & \textbf{HRA@VNI} \\
\midrule
\textbf{email-Enron} &
\lightPurple{0.2542$\pm$0.0444} &
\deepPurple{0.2608$\pm$0.0433} &
0.2530$\pm$0.0444 &
0.2537$\pm$0.0440 &
0.2296$\pm$0.0416 &
0.2530$\pm$0.0408 \\
\textbf{email-Eu} &
0.2321$\pm$0.0178 &
\deepPurple{0.4312$\pm$0.0152} &
0.3009$\pm$0.0180 &
0.3023$\pm$0.0179 &
0.2921$\pm$0.0166 &
\lightPurple{0.3917$\pm$0.0153} \\
\textbf{DAWN} &
0.1701$\pm$0.0107 &
0.2083$\pm$0.0107 &
0.1898$\pm$0.0108 &
0.1955$\pm$0.0108 &
\lightPurple{0.2411$\pm$0.0114} &
\deepPurple{0.2535$\pm$0.0110} \\
\textbf{NDC-classes} &
-0.0936$\pm$0.0096 &
\lightPurple{0.3342$\pm$0.0087} &
0.1947$\pm$0.0086 &
0.1731$\pm$0.0085 &
0.3174$\pm$0.0090 &
\deepPurple{0.3430$\pm$0.0082} \\
\textbf{congress-bills} &
\deepPurple{0.4324$\pm$0.0220} &
\lightPurple{0.4269$\pm$0.0203} &
0.4205$\pm$0.0220 &
0.4212$\pm$0.0220 &
0.2169$\pm$0.0200 &
0.3139$\pm$0.0189 \\
\textbf{contact-high-school} &
0.2877$\pm$0.0263 &
\lightPurple{0.3434$\pm$0.0278} &
0.2784$\pm$0.0266 &
0.2769$\pm$0.0258 &
0.3392$\pm$0.0249 &
\deepPurple{0.3457$\pm$0.0255} \\
\textbf{contact-primary-school} &
0.1223$\pm$0.0385 &
\lightPurple{0.1564$\pm$0.0385} &
0.0998$\pm$0.0397 &
0.0964$\pm$0.0395 &
0.1868$\pm$0.0387 &
\deepPurple{0.1890$\pm$0.0388} \\
\textbf{tags-ask-ubuntu} &
0.2950$\pm$0.0115 &
0.3251$\pm$0.0115 &
0.2967$\pm$0.0115 &
0.3140$\pm$0.0118 &
\lightPurple{0.3993$\pm$0.0124} &
\deepPurple{0.4025$\pm$0.0112} \\
\textbf{tags-math-sx} &
0.4140$\pm$0.0169 &
\lightPurple{0.4354$\pm$0.0167} &
0.4131$\pm$0.0169 &
0.4290$\pm$0.0167 &
0.4103$\pm$0.0172 &
\deepPurple{0.4482$\pm$0.0164} \\
\textbf{iAF1260b} &
0.0798$\pm$0.0108 &
0.0833$\pm$0.0109 &
0.0892$\pm$0.0107 &
0.0820$\pm$0.0107 &
0.0414$\pm$0.0132 &
\deepPurple{0.1149$\pm$0.0116} \\
\textbf{iAF987} &
0.1896$\pm$0.0138 &
0.1902$\pm$0.0139 &
\lightPurple{0.1903$\pm$0.0137} &
0.1861$\pm$0.0140 &
0.0245$\pm$0.0139 &
\deepPurple{0.2013$\pm$0.0133} \\
\textbf{iCN900} &
0.2010$\pm$0.0130 &
0.2001$\pm$0.0127 &
\lightPurple{0.2012$\pm$0.0129} &
0.1955$\pm$0.0128 &
0.2009$\pm$0.0140 &
\deepPurple{0.2221$\pm$0.0118} \\
\bottomrule
\end{tabular}
}
\end{table}

To evaluate the effectiveness of the proposed method in real-world propagation scenarios, we adopt the nonlinear SIR model on hypergraphs~\cite{St-Onge2022,hu2023}, which can characterize the nonlinear spreading mechanism induced by group effects in high-order interactions. Nodes in this model can be in one of three possible states: susceptible (S), infected (I) and recovered (R). A hyperedge is called infected hyperedge if it contains at least one infected node. This nonlinear SIR process begins with one or more infected seeds and all other nodes are initially susceptible. At each time step, if a susceptible node $v_i$ belongs to an infected hyperedge $e_\alpha$ that contains $\eta_{i\alpha}$ infected nodes, $v_i$ will be infected by $e_\alpha$ with a probability $\beta \eta_{i\alpha}^{\kappa}$. Thus, the probability that node $v_i$ will be infected at this time step is
\begin{equation}
p_i^{S\rightarrow I}=1-\prod_\alpha\left( 1-\beta \eta_{i\alpha}^{\kappa} \right).
\end{equation}
At the same time, each infected node will turn to be a recovered node with probability $\gamma$. Generally speaking, the choice of parameter $\gamma$ does not affect the essence of the spreading dynamics, since the influence caused by variations in $\gamma$ can be offset by adjusting $\beta$. For this reason, previous studies on the SIR model usually set $\gamma=1$ \cite{zhou2006,pastor2015,wang2017}. We have noted that a smaller $\gamma$ tends to yield more refined simulation results and reduce fluctuations, but it also leads to an increase in simulation time. To strike a balance, we set $\gamma=0.25$ in this work. Existing studies on this nonlinear SIR model mainly focus on the case where $\kappa>1$~\cite{St-Onge2022,hu2023}, because $\kappa>1$ can characterize the social reinforcement effect in information propagation~\cite{centola2010,lu2011b}. Therefore, we report the results for $\kappa=1.25$ in the main text and the result for more $\kappa$-values in the SI Appendix.

For a real hypergraph $G$, the influence of a node $v_i$, say $I_i$, is defined as the number of eventually recovered nodes averaged over a certain number of independent runs, each of which starts with node $v_i$ being the sole infected seed. In this study, we always use 100 independent runs. Given the target centrality measure $C$ (e.g., $c^{\mathrm{HRA}}$) and the corresponding centrality values $c_1, c_2, \cdots, c_N$ of all nodes, we quantify to what extent the measure $C$ resembles spreading influences of individual nodes by calculating the rank correlation between centrality values $c_1, c_2, \cdots, c_N$ and node influences $I_1, I_2, \cdots, I_N$, using the Kendall's $\tau$ \cite{Kendall1938}. Considering any two vectors associated with all $N$ nodes, $X=(x_1,x_2,\cdots,x_N)$ and $Y=(y_1,y_2,\cdots,y_N)$, as well as the $N$ two-tuples $(x_1,y_1),(x_2,y_2),\cdots,(x_N,y_N)$, then, any pair $(x_i,y_i)$ and $(x_j,y_j)$ are concordant if the ranks for both elements agree, namely if both $x_i>x_j$ and $y_i>y_j$ or if both $x_i<x_j$ and $y_i<y_j$. They are discordant if $x_i>x_j$ and $y_i<y_j$ or if $x_i<x_j$ and $y_i>y_j$. If $x_i=x_j$ or $y_i=y_j$, the pair is neither concordant or discordant. Comparing all $N(N-1)/2$ pairs of two-tuples, the Kendall's $\tau$ is defined as
\begin{equation}
	\tau(X,Y)=\frac{2(n_+-n_-)}{N(N-1)} \in [-1, 1],
\end{equation}
where $n_+$ and $n_-$ are the number of concordant and discordant pairs, respectively. If $X$ and $Y$ are independent, $\tau$ should be close to zero, and thus the extent to which $\tau$ exceeds zero indicates the strength of correlation. The subsequent experiments mainly focus on comparing $\tau(c^{\mathrm{HRA}},I)$ with other benchmark methods. 

\begin{figure}[htbp]
    \centering
    \includegraphics[width=1.0\columnwidth]{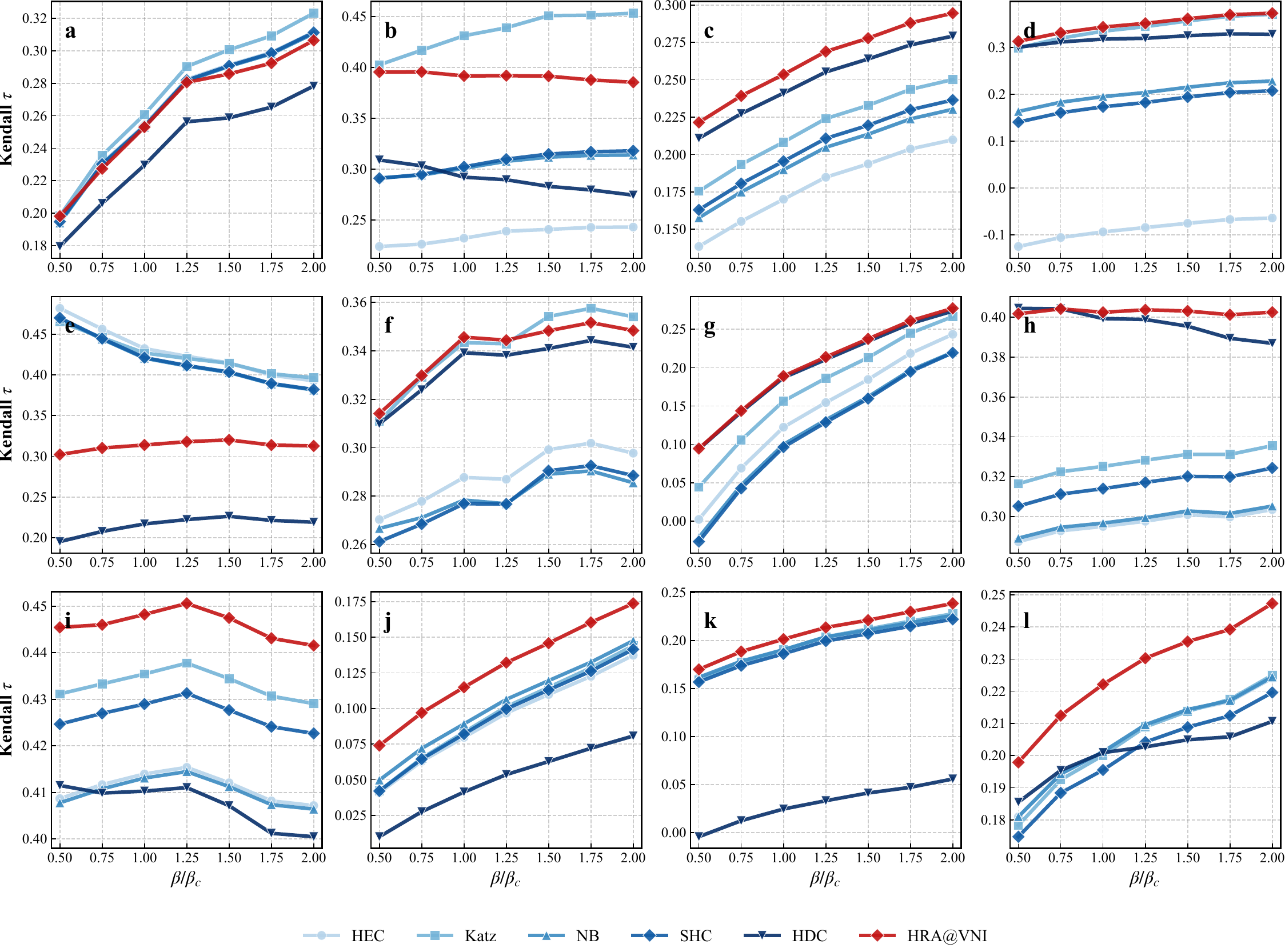}
    \caption{\label{f_Kendall_tau} {Comparative results of the performance of methods to identify vital nodes. The x-axis denotes the infectivity $\beta$, scaled as $\beta/\beta_c$, the y-axis is the corresponding kendall's $\tau$, obtained by averaging over 100 independent runs. The results of different algorithms are represented by curves of different colors, among which the algorithm HRA@VNI proposed in this paper is marked in red.}}
\end{figure}

Notice that, the choice of $\beta$ is nontrivial. When $\beta$ is very small, the disease cannot spread out and the infected node only has a small chance to infect its immediate neighbors, so that the problem to estimate a node's spreading influence becomes trivial and the best index is just the number of immediate neighbors, say degree. In contrast, when $\beta$ is very high, the disease will infect a large percentage of the population, irrespective of where it originated, and thus the individual influence is meaningless. Accordingly, we focus on the case of $\beta=\beta_c$, where $\beta_c$ is the epidemic threshold \cite{chakrabarti2008,castellano2010}. Although there exists well‑established methods to analytically derive the threshold for the SIR model on simple networks \cite{pastor2015,wang2017,chakrabarti2008,castellano2010}, the nonlinear SIR model on hypergraphs is considerably more complex in both topology and dynamics. Consequently, simple analytical expressions are unavailable, and we thus employ numerical simulations to determine the existence and value of the epidemic threshold (see Figure S1 in SI Appendix). For intuitive illustration, we plot the curve of $R(t)$ versus time step $t$ for $\beta=\beta_c$ (see Figure S2 in SI Appendix), where $R(t)$ denotes the number of recovered nodes after time step $t$.

Table 4 compares the performance of the proposed HRA@VNI method with five representative benchmark algorithms for vital nodes identification in hypergraphs under the condition 
$\beta=\beta_c$. These five benchmarks are hypergraph eigenvector centrality (HEC) \cite{tudisco2021}, Katz Centrality (Katz) \cite{katz1953}, nonbacktracking centrality (NB) \cite{krzakala2013,martin2014,gautier2019}, subhypergraph centrality (SHC) \cite{estrada2006}, and hypergraph degree centrality (HDC) \cite{battiston2020} (see Methods for algorithm details). As shown in Table 4, the HRA@VNI method achieves the best performance on 9 out of 12 real-world hypergraphs, demonstrating a significant advantage over other algorithms. We further consider a wider range of values for $\beta$ near the epidemic threshold. Figure 2 presents the comparative results of the six algorithms under the condition $\beta \in [0.5\beta_c, 2\beta_c]$. It can be seen that these results are consistent with those in Table 4: in the vast majority of considered hypergraphs, the HRA@VNI method shows a consistent and significant advantage. Figures S3-S5 (see Section S3 in SI Appendix) show the results for more $\kappa$-values (Fig. S3 for $\kappa=1$, Fig. S4 for $\kappa=1.5$, and Fig. S5 for $\kappa=2$), where the HRA@VNI method consistently exhibits a significant advantage compared with other benchmark algorithms. Given that HRA@VNI is merely an extremely simple direct application based on $\mathcal{P}$, the above results strongly support the rationality and application potential of the proposed proximity matrix.

\section{Discussion}
Higher-order interactions are the key structural ingredients underlying the emergence of macroscopic ordered behaviors and the formation of statistical regularities in complex systems. Since the study of hypergraphs became a hot topic in network science, the binary incidence matrix $\mathcal{M}$ has served as the cornerstone for characterizing and analyzing hypergraphs. Although $\mathcal{M}$ can represent a hypergraph without information loss, it only encodes the membership relation and discards the strength information. Due to the heterogeneity of node degrees and hyperedge orders, the membership relation is far from sufficient to characterize the actual participation level of nodes in collective processes. The proximity matrix $\mathcal{P}$ proposed in this study is precisely intended to remedy this representational deficiency. Through a resource allocation process, $\mathcal{P}$ supplements the binary topological relations with continuous strength information. 

To verify the effectiveness of the proposed proximity matrix $\mathcal{P}$, we evaluated the performance of algorithms based on $\mathcal{P}$ on three mainstream tasks in hypergraph mining: link prediction, community detection, and vital nodes identification. Surprisingly, simple algorithms centered on $\mathcal{P}$ can achieve significantly better performance than benchmark methods across all three tasks. The superior performance of the proposed algorithms can be attributed to the fact that $\mathcal{P}$ can well capture the heterogeneity of hypergraph structures and thus accurately quantify the relationships between nodes and hyperedges. As the incidence matrix $\mathcal{M}$ is the classical and most commonly used way to characterize node-hyperedge relationship, we compare the performance of the binary $\mathcal{M}$ matrix and the continuous-valued $\mathcal{P}$ matrix on the three hypergraph mining tasks by directly replacing $\mathcal{P}$ with $\mathcal{M}$ in the three proposed algorithms (HRA@LP, HRA@CD, and HRA@VNI) and juxtaposing the two types of algorithms before and after the replacement. As shown in Table S3 (see Section S4 in SI Appendix), $\mathcal{P}$-based algorithms outperforms $\mathcal{M}$-based counterparts in most cases, further validating the effectiveness of $\mathcal{P}$.

Notice that, the proximity matrix $\mathcal{P}$ is obtained by applying the resource allocation operator $\mathcal{T}$ to the incidence matrix $\mathcal{M}$. If we set $\mathcal{P}^{(0)} \equiv \mathcal{M}$ and $\mathcal{P}^{(t)}=\mathcal{T} \cdot \mathcal{P}^{(t-1)}$, then we can get a sequence of matrices $\mathcal{P}^{(0)} \equiv \mathcal{M}, \mathcal{P}^{(1)} \equiv \mathcal{P}, \mathcal{P}^{(2)},\mathcal{P}^{(3)},\cdots$. As $t$ increases, the sequence $\mathcal{P}^{(t)}$ converges to $\mathcal{P}^{(\infty)}$, whose entries are given by $\mathcal{P}^{(\infty)}_{i\alpha}=\frac{d_ik_\alpha}{\sum_j d_j}$ (see Section S5 in SI Appendix for the mathematical proof). Although each matrix $\mathcal{P}^{(t)}$ in this sequence can serve as a proximity matrix, considering very large $t$ is of limited value, since it increases computational cost, reduces interpretability, and dilutes important local information as $t$ increases. In fact, even relatively small $t$ ($t>1$) can be harmful, as it may create connections between a large number of nodes and hyperedges, thereby degrading the value of effective information \cite{zhou2009b}. Of course, we do not rule out the possibility that choosing certain small values of $t$ in some networks may moderately improve performance on specific hypergraph mining tasks \cite{wu2014}. Empirical studies in this direction can be conducted in the future.

By replacing the binary matrix $\mathcal{M}$ with the continuous-valued matrix $\mathcal{P}$, we achieve significant performance improvements across three distinct hypergraph mining tasks. This demonstrates that a better representation of node‑hyperedge relationships has the potential to universally enhance hypergraph analysis. It is analogous to how a superior graph embedding algorithm can simultaneously boost the performance of numerous graph-based algorithms \cite{goyal2018,cai2018}. Similarly, the proximity matrix $\mathcal{P}$ is not limited to the three hypergraph mining tasks addressed here; it can serve as a fundamental tool for various aspects of hypergraph analysis. For instance, the Ollivier-Ricci curvature of hypergraphs defined based on the $\mathcal{M}$ matrix fails to sensitively capture the genuine fluctuations in the intensity of local many-body interactions \cite{Eidi2020, coupette2023}, which may lead to degraded resolution in optimal transport problems. In contrast, adopting the $\mathcal{P}$ matrix has the potential to improve such resolution. Analogously, the continuous-valued $\mathcal{P}$ matrix can characterize local curvature fluctuations under the Wasserstein distance \cite{Asoodeh2018}, thereby providing analyses with rigorous differential geometric meaning for identifying geometric constraints on information propagation in hypergraphs. In addition to $\mathcal{M}$, another widely used matrix for representing hypergraphs is the Laplacian matrix $\mathcal{L}$, which is also defined based on $\mathcal{M}$ \cite{chung1993,nurisso2025}. For example, one of the most common forms is $\mathcal{L}=\mathcal{D}-\mathcal{M}\mathcal{K}^{-1}\mathcal{M}^\top$. If we replace the incidence matrix $\mathcal{M}$ with the proximity matrix $\mathcal{P}$  or other matrices serving a similar role (with necessary adjustments to the normalization method), we can obtain a series of new Laplacian matrices. Since dynamical processes of hypergraphs are often characterized by Laplacian matrices \cite{Perc2013,Carletti2020,Skardal2020,Millan2020,Alvarez-Rodriguez2021,Majhi2022,Wang2024,Arruda2024}, those new Laplacian matrices correspond to new weighted dynamical processes. Studying such weighted dynamics may improve the accuracy in characterizing hypergraph functions and advance the discovery of new dynamical properties of hypergraphs \cite{Chitra2019,Pan2025}.

\section{Methods}

This section will introduce all benchmark algorithms for the three considered hypergraph mining tasks. 

\subsection{Link Prediction Benchmarks}

We consider four benchmark hypergraph link prediction algorithms based on distinct mechanisms, including neighborhood overlap, resource allocation, global path aggregation, and graph embedding. Specifically, they are common neighbors index (CN) \cite{liben2007, zhang2018}, hyperedge prediction using resource allocation (HPRA) \cite{kumar2020}, Katz index (Katz) \cite{katz1953, zhang2018}, and nonuniform hypergraph embedding with dual mechanism (NHNE) \cite{huang2020}.

\textbf{CN} \cite{liben2007, zhang2018} is a local similarity heuristic that quantifies the likelihood of a connection between two nodes by counting their shared neighbors. For nodes $v_i$ and $v_j$, the CN index is defined as
\begin{equation}
S^{CN}_{ij} = |\Gamma(v_i) \cap \Gamma(v_j)|,
\end{equation}
where $\Gamma(v_i)$ denotes the set of nodes that co-occur with $v_i$ in at least one hyperedge. Similar to the HRA@LP method, the score of a node set (i.e., a candidate hyperedge) is set as the average similarity over all node pairs in this set.

\textbf{HPRA} \cite{kumar2020} extends the resource allocation index to hypergraphs by explicitly accounting for node co-membership within hyperedges. The direct contribution to the HPRA index for any pair of nodes $v_i$ and $v_j$ is 
\begin{equation}
\tilde{S}^{HPRA}_{ij} = \sum_{e_{\alpha}: v_i,v_j \in e_{\alpha}} \frac{1}{k_{\alpha}-1},
\end{equation}
where $e_{\alpha}$ denotes a hyperedge containing both $v_i$ and $v_j$, and $k_{\alpha}$ is the order of $e_{\alpha}$. Indirect contributions mediated by common neighbors of $v_i$ and $v_j$ are further incorporated, leading to the complete definition of HPRA index
\begin{equation}
S^{HPRA}_{ij} = \tilde{S}^{HPRA}_{ij} + \sum_{v_l \in \Gamma(v_i) \cap \Gamma(v_j)} \frac{1}{d_l} \tilde{S}^{HPRA}_{il} \tilde{S}^{HPRA}_{lj},
\end{equation}
where $d_l$ denotes the degree of node $v_l$. The score of a candidate hyperedge is the average of HPRA indices over all node pairs in this candidate.

\textbf{Katz} \cite{katz1953} evaluates node similarity by aggregating contributions from all possible paths connecting them, with longer paths exponentially attenuated. In the hypergraph setting, Katz index is computed on the hypergraph adjacency matrix $\mathcal{A} = \mathcal{M}\mathcal{M}^{\top} - \mathcal{D}$ as
\begin{equation}
S^{Katz}_{ij} = \sum_{l=1}^{\infty} \lambda^l (\mathcal{A}^l)_{ij}
          = [(\mathcal{I} - \lambda \mathcal{A})^{-1} - \mathcal{I}]_{ij},
\end{equation}
where $\lambda$ is the attenuation factor controlling the contribution of longer paths. Analogously, the score of a candidate hyperedge is the average of Katz indices over all node pairs in this candidate.

\textbf{NHNE} \cite{huang2020} is a learning-based framework designed to capture higher-order and nonuniform relationships through a dual-mechanism architecture. Any hyperedge $e$, whether it is in the training set or designated as a candidate for prediction, is represented using a position-based one-hot encoding, say $\mathcal{X}_e \in \mathbb{R}^{N \times z}$, where $z$ is the maximum hyperedge order. For hyperedges whose order is smaller than $z$, zero padding is applied. Each column of $\mathcal{X}_e$ corresponds to a node-position indicator within the hyperedge. This encoding is then multiplied by the embedding layer weights $\mathcal{W}^{(emb)}$ to obtain the initial input representation. Subsequently, two parallel one‑dimensional convolutional networks (1D‑CNNs) are applied to extract hyperedge features from the original and dual hypergraphs, respectively. The outputs of these networks are concatenated and passed through a fully connected layer to produce the similarity score of a hyperedge, defined as
\begin{equation}
    S(e) = \sigma(\mathcal{W}^{(out)} [c_o(\mathcal{W}_o^{(emb)} \mathcal{X}_e);c_d(\mathcal{W}_d^{(emb)}\mathcal{X}_e)] + b),
\end{equation}
where $\sigma$ denotes the sigmoid activation function, $\mathcal{W}_o^{(emb)}$ and  $\mathcal{W}_d^{(emb)}$ are the embedding weights for the original and dual hypergraphs, $c_o$ and $c_d$ correspond to the respective 1D‑CNN feature extractors, $\mathcal{W}^{(out)}$ is the weight matrix of the output layer, and $b$ is the bias term.

\subsection{Community Detection Benchmarks}

For the community detection task, we compare the HRA@CD method with five representative algorithms that are based on distinct principles, including matrix factorization, modularity maximization, graph partitioning, nonbacktracking spectral analysis, and hypergraph Laplacian method. Specifically, The five benchmark methods are non-negative matrix factorization (NMF) \cite{lee2000}, node degree preserving Louvian algorithm (NDP-Louvian) \cite{kumar2020b}, Metis algorithm for hypergraph adjacency matrix (AMetis) \cite{karypis1998, catalyurek1999}, hypergraph spectral clustering (HSC) \cite{zhouD2006}, and nonbacktracking hypergraph spectral clustering (NBHSC) \cite{chodrow2023}. 

\textbf{NMF} \cite{lee2000} identifies community structure by factorizing the hypergraph adjacency matrix into low-rank nonnegative components. Given the adjacency matrix $\mathcal{A} \in \mathbb{R}^{N \times N}$, NMF seeks a decomposition
\begin{equation}
\mathcal{A} \approx \mathcal{F}\mathcal{Z},
\end{equation}
where $\mathcal{F} \in \mathbb{R}^{N \times N_c}$ represents node-to-community membership strengths and $\mathcal{Z} \in \mathbb{R}^{N_c \times N}$ encodes inter-community connectivity, with all entries constrained to be nonnegative.
The factorization is obtained by minimizing the reconstruction error
\begin{equation}
\min_{\mathcal{F},\mathcal{Z} \ge 0} \|\mathcal{A} - \mathcal{F}\mathcal{Z}\|_F^2.
\end{equation}
After convergence, node $v_i$ is assigned to the community corresponding to its largest membership value in $\mathcal{F}$, say
\begin{equation}
g_i=\arg\max_a \mathcal{F}_{ia},
\end{equation}
where $g_i$ denotes the community assignment of node $v_i$.

\textbf{NDP-Louvain} \cite{kumar2020b} detects communities by maximizing a degree-preserving modularity defined on hypergraphs. In the null hypergraph, the expected number of edges between nodes $v_i$ and $v_j$ is
\begin{equation}
\mathcal{E}_{ij} = \frac{d_i d_j}{\sum_{v_l \in V} d_l}.
\end{equation}
To enable modularity optimization, the target hypergraph is first mapped to a weighted one via
\begin{equation}
\mathcal{A}^W = \mathcal{M}(\mathcal{K}-\mathcal{I})^{-1}\mathcal{M}^\top,
\end{equation}
the modularity matrix is defined as
\begin{equation}
\mathcal{B}_{ij} = \mathcal{A}_{ij}^W - \mathcal{E}_{ij},
\end{equation}
and the corresponding modularity objective is
\begin{equation}
Q = \frac{1}{2m} \sum_{ij} \mathcal{B}_{ij} \delta(g_i, g_j).
\end{equation}
This objective is optimized using the Louvain algorithm. Since the Louvain algorithm does not allow direct control over the number of detected communities, we apply an additional agglomerative clustering step based on average linkage \cite{sokal1958} to merge the communities obtained by the Louvain algorithm until the number of communities reaches $N_c$.

\textbf{AMetis} \cite{karypis1998, catalyurek1999} performs hypergraph partitioning by first transforming the hypergraph into a weighted graph via clique reduction. The resulting weighted adjacency matrix is then partitioned using the Metis algorithm.

\textbf{HSC} \cite{zhouD2006} extends spectral clustering to hypergraphs through a normalized hypergraph Laplacian. From the incidence matrix $\mathcal{M}$, a weighted adjacency matrix is constructed as
\begin{equation}
\mathcal{A}^{Z} = \mathcal{M}\mathcal{K}^{-1}\mathcal{M}^\top,
\end{equation}
the normalized Laplacian is then defined as
\begin{equation}
\mathcal{L}^{Z} = \mathcal{I} - \mathcal{D}^{-1/2}\mathcal{A}^{Z}\mathcal{D}^{-1/2}.
\end{equation}
The eigenvectors corresponding to the smallest eigenvalues of $\mathcal{L}^{Z}$ are used as node embeddings, followed by $k$-means clustering algorithm with $k$ is set to be $N_c$.

\textbf{NBHSC} \cite{chodrow2023} is a spectral community detection method based on a hypergraph nonbacktracking operator. We define $(v_i,e_\alpha)$ as a node-hyperedge incidence pair if $v_i \in e_\alpha$, and then a transition from $(v_i,e_\alpha)$ to $(v_j,e_\beta)$ is allowed only if $v_j \in e_\alpha$, $e_\beta \neq e_\alpha$, and $v_j \neq v_i$, thereby preventing immediate reversal along the same local structure as $v_i$ generally does not belong to $e_\beta$. Since the direct eigendecomposition of the full nonbacktracking operator is expensive, NBHSC instead works with a reduced matrix $\mathcal{B}'$, derived from an Ihara--Bass relation for nonuniform hypergraphs, as
\begin{equation}
\mathcal{B}' =
\begin{pmatrix}
0_{\lvert K \rvert N} & \mathcal{V} - \mathcal{I}_{\lvert K \rvert N} \\
(\mathcal{I}_{\lvert K \rvert} - \mathcal{Y}) \otimes \mathcal{I}_{N} &
\mathcal{U} + (2\mathcal{I}_{\lvert K \rvert} - \mathcal{Y}) \otimes \mathcal{I}_{N}
\end{pmatrix},
\end{equation}
where $\mathcal{U}$ and $\mathcal{V}$ denote the block matrices collecting the order-specific adjacency and degree operators, respectively, $\mathcal{Y}$ is a diagonal matrix whose diagonal entries are the possible hyperedge orders, $K$ is the set of possible hyperedge orders, $|K|$ is its cardinality, and $\otimes$ denotes the Kronecker product.

Given $N_c$, NBHSC extracts the eigenvectors of $\mathcal{B}'$ corresponding to the $N_c$ largest eigenvalues and discards the leading uninformative one, leaving $N_c-1$ informative spectral directions. For each retained eigenvector, NBHSC aggregates the entries of its second block associated with node $v_i$ across hyperedge-order blocks to obtain
\begin{equation}
\bar{x}_i^{(\ell)}=\sum_{r \in K} x_{2;r,i}^{(\ell)},
\end{equation}
where $x_{2;r,i}^{(\ell)}$ is the entry corresponding to node $v_i$ in the block for hyperedge order $r$, and $\ell$ indexes the retained nontrivial eigenvectors. The aggregated score is then binarized as
\begin{equation}
\tilde{x}_i^{(\ell)}=\operatorname{sgn}\!\bigl(\bar{x}_i^{(\ell)}\bigr),
\end{equation}
and the binarized vectors from all retained eigenvectors are stacked into a Euclidean node embedding, on which $k$-means is applied with the number of clusters set to $N_c$.

\subsection{Vital Nodes Identification Benchmarks}

To evaluate the effectiveness of the proposed method in identifying vital nodes in hypergraphs, we compare HRA@VNI with five representative centrality measures that characterize node importance from different structural perspectives. These include hypergraph eigenvector centrality (HEC) \cite{tudisco2021}, Katz Centrality (Katz) \cite{katz1953}, nonbacktracking centrality (NB) \cite{krzakala2013,martin2014,gautier2019}, subhypergraph centrality (SHC) \cite{estrada2006}, and hypergraph degree centrality (HDC) \cite{battiston2020}

\textbf{HEC} \cite{tudisco2021} models node influence through iterative refinement between nodes and hyperedges: the importance of a node depends on the importance of the hyperedges it participates in, while the importance of a hyperedge depends on the centrality of its incident nodes. The node centrality vector $x\in\mathbb{R}^N$ and hyperedge centrality vector $y\in\mathbb{R}^M$ satisfy
\begin{equation}
\begin{cases}
\lambda x = \mathcal{MK}y \\
\mu y = \mathcal{M}^\top \mathcal{D}x
\end{cases}
\quad \lambda, \mu > 0.
\end{equation}
The above coupled system is solved via power iteration
\begin{equation}
   x^{(t+1)} = \mathcal{M}\mathcal{K} y^{(t)}, 
   \qquad
   y^{(t+1)} = \mathcal{M}^\top \mathcal{D} x^{(t+1)}, 
\end{equation}
followed by normalization of $x^{(t+1)}$ and $y^{(t+1)}$ at each step. The vectors are initialized as uniform positive vectors, i.e., $x^{(0)}=\frac{1}{N}\mathbf{1}_N$ and $y^{(0)}=\frac{1}{M}\mathbf{1}_M$. The scaling factors $\lambda$ and $\mu$ correspond to the dominant eigenvalue factors and are implicitly determined during normalization.

\textbf{Katz} \cite{katz1953} measures node influence by aggregating contributions from all possible walks in the target hypergraph, with longer walks exponentially attenuated. The Katz centrality score of node $v_i$ is defined as
\begin{equation}
x_i = 1+ \gamma \sum_{j} \mathcal{A}_{ij} x_j ,
\end{equation}
where $\mathcal{A}$ is the hypergraph adjacency matrix, and $\gamma$ is the attenuation factor.

\textbf{NB} \cite{krzakala2013,martin2014,gautier2019} characterizes node influence by suppressing backtracking paths that immediately return to the previous node, thereby highlighting genuine spreading capability. In hypergraphs, nodes and hyperedges are treated as the two parts of a bipartite graph, and the nonbacktracking propagation matrix is defined as
\begin{equation}
\mathcal{N} =
\begin{bmatrix}
0 & \mathcal{M} \\
\mathcal{M}^\top & 0
\end{bmatrix}
\in \mathbb{R}^{(N+M)\times(N+M)}.
\end{equation}
If $\mathcal{N}$ is nonnegative and irreducible, the generalized Perron--Frobenius theorem guarantees the existence of a unique positive principal eigenvector. The nonbacktracking centrality of node $v_i$ is then given by the $i$-th entry of this eigenvector, denoted by $c_i^{\mathrm{NB}}$.

\textbf{SHC} \cite{estrada2006} quantifies node importance by counting closed walks of all lengths that start and end at the target node, reflecting its participation in global substructures. It is defined via the spectral decomposition of the hypergraph adjacency matrix $\mathcal{A}$ as
\begin{equation}
c^{\mathrm{SHC}}_i = \sum_{j=1}^{N} \xi_{ij}^2 \exp(\lambda_j),
\end{equation}
where $\lambda_j$ is the $j$-th eigenvalue of $\mathcal{A}$ and $\xi_{ij}$ is the $i$-th entry of the corresponding eigenvector.

\textbf{HDC} \cite{battiston2020} is adopted as a baseline measure of node importance in hypergraphs, defind as 
\begin{equation}
c^{\mathrm{HDC}}_i = d_i.
\end{equation}

\FloatBarrier

\end{document}